\documentclass[runningheads]{llncs}
\usepackage[T1]{fontenc}
\usepackage{graphicx}
\usepackage{amsmath}
\usepackage{amsfonts}
\usepackage{amssymb}
\usepackage{multirow}
\usepackage{ifthen}
\usepackage{doi}
\usepackage{physics}

\usepackage[
backend=biber,
style=numeric,
sorting=none
]{biblatex}
\usepackage{xcolor}
\graphicspath{{./Figs/}}
\newcommand{\bassem}[1]{#1}

\begin{document}
\title{General control of linear cellular automata}
%
\author{Franco Bagnoli\inst{1,2,4\dagger}\orcidID{0000-0002-6293-0305}  \and
Sara Dridi\inst{3}\orcidID{0000-0002-1212-2246} \and
Bassem Sellami\inst{2}\orcidID{0000-0001-6869-3518} \and
Amira Mouakher\inst{2}\orcidID{0000-0002-1346-3851}\and
Samira El Yacoubi\inst{2}\orcidID{0000-0002-8017-5286}
}
\authorrunning{F. Bagnoli et al.}

\institute{Department of Physics and Astronomy and CSDC, University of Florence,  via G. Sansone 1, 50019 Sesto Fiorentino (Italy) \email{franco.bagnoli@unifi.it}, 
\and Espace-Dev, UPVD, IRD, UM, Perpignan (France) \email{\{yacoubi,bassem.sellami, amira.mouakher\}@univ-perp.fr}
\and Institute of Optics and Precision Mechanics,  University of Setif 1 (Algeria) \email{sara.dridi@univ-setif.dz}
\and INFN, Sect. Florence (Italy) 
}
\maketitle             

\let\at@
\catcode`@=\active
\def@#1{\ifmmode\boldsymbol{#1}\else\at#1\fi}

\let\quot"
\catcode`"=\active
\def"#1"{``#1''}

\renewcommand\P{\mathcal{P}}
\renewcommand\H{\mathcal{H}}
\newcommand\Z{\mathcal{Z}}

\newcommand{\eq}[2][]{%
    \ifthenelse{\equal{#1}{}}{%
        \begin{equation*}
            #2%
        \end{equation*}%
    }{%
        \begin{equation}\label{eq:#1}%
            #2%
        \end{equation}%
    }%
}
\newcommand{\meq}[2][]{%
    \ifthenelse{\equal{#1}{}}{%
        \begin{equation*}%
            \begin{split}%
                #2%
            \end{split}%
        \end{equation*}%
    }{%
        \begin{equation}\label{eq:#1}%
            \begin{split}%
                #2%
            \end{split}%
        \end{equation}%
    }%
}
\newcommand{\Eq}[1]{Eq.~\eqref{eq:#1}}

\newcommand{\eps}{\varepsilon}

\newcommand{\A}{\mathrm{A}}
\newcommand{\B}{\mathrm{B}}
\newcommand{\C}{\mathrm{C}}
\newcommand{\D}{\mathrm{D}}
\newcommand{\E}{\mathrm{E}}
\newcommand{\F}{\mathrm{F}}

\begin{abstract}
In mathematics and engineering, control theory is concerned with the analysis of dynamical systems through the application of suitable control inputs. One of the prominent  problems in control theory is controllability which concerns the ability to determine whether there exists a control input that can steer  a dynamical system from  an initial state to a desired final state within a finite time horizon. There is a general theory for controlling linear or linearizable system, but it cannot be applied to discrete systems like cellular automata, which is the problem of that we address in this paper. We develop a general theory for linear (and affine) cellular automata, and apply it to examples of one-dimensional and two-dimensional Boolean cases. We introduce the concept of controllability matrix and show that controllability holds if and only if the controllability matrix is invertible.
\end{abstract}


\section{Introduction}\label{sec:introduction}
 
Control theory is a field of applied mathematics that forms a strong link between mathematical theory and technological applications. It is
concerned with the behavior of dynamical systems and the ways in which this behavior can be influenced through external inputs. 

Among the fundamental concepts in control theory there is the notion of controllability, which has played a crucial role in the development of modern control theory and in the resolution of many control problems. It refers to the ability to design appropriate control inputs that steer a dynamical system from a given initial state to a specified target state within a specified time interval. Since its introduction, this concept has been examined for infinite dimensional systems described by partial differential equations \cite{lions1971optimal,Lions1988} and finite dimensional systems \cite{sontag2013mathematical} \bassem{as well as for finite-dimensional systems governed by linear or nonlinear differential or difference equations~\cite{Kalman1960}. More recently, controllability has also been studied for cellular automata~\cite{dridi2019markov,dridi2019recent,dridi2019graph,dridi2022kalman,bagnoli2016toward,bagnoli2018regional,bagnoli2019optimal,Bagnoli2024,dridi2019boundary,beros2019controlled}.}
\bassem{For linear or linearizable systems, a general control theory can be built using standard tools such as Fourier or Laplace transforms, and a central difficulty is often the reconstruction of the system state from noisy measurements~\cite{Kalman1960}.} More recently, it has been studied within the formwork of cellular automata \cite{dridi2019markov,dridi2019recent,dridi2019graph,dridi2022kalman,bagnoli2016toward,bagnoli2018regional,bagnoli2019optimal,Bagnoli2024,dridi2019boundary,beros2019controlled}.

A general control theory can be developed for linear or linearizable systems, i.e., systems described by a system of linear differential equations. 
In this case one can apply techniques based on suitable transformations of the system (Fourier, Laplace etc.), and the difficulties are mainly that of identifying the state of the system from noisy  measurements~\cite{Kalman1960}. 
In the case of discrete systems like cellular automata, the problem is quite different. Due to their discrete nature, there is less concern about noise in measurements, but they need a completely different approach, even for ``linear'' ones, as we shall see in the following. \bassem{Indeed, CA evolve over finite alphabets through local update rules, so classical tools based on linear algebra over the reals or on Fourier and Laplace transforms do not apply directly, even in the linear case. This calls for a framework adapted to both locality and Boolean structure.}

Cellular automata (CA) are discrete dynamical systems that offer a good tool  for modeling and analyzing natural phenomena. They are defined by three fundamental components: a grid of cells or sites, each taking a state from a finite set, a specified neighborhood for every cell, and a local transition function that determines how each cell updates its state based on the states within its neighborhood. This transition function can be deterministic or stochastic. Cellular automata, beyond being an interesting subject of mathematics and theoretical physics,  have been applied to the modeling in various fields, including  physics, biology, and chemistry~\cite{ACRI}. 
\bassem{Recent studies have addressed controllability in CA through boundary control, regional control, and Markovian representations of the dynamics~\cite{dridi2019markov,dridi2019recent,dridi2019graph,dridi2022kalman,bagnoli2016toward,bagnoli2018regional,bagnoli2019optimal,Bagnoli2024,dridi2019boundary,beros2019controlled}. However, a general and constructive finite-time control theory, comparable in spirit to the Kalman framework, is still lacking.}
In this paper we deal with the general control theory of linear and affine deterministic cellular automata, defined as the ability to  drive their evolution from a given initial configuration to a desired final configuration,
by modifying the values of a minimal number of sites in given positions and at different time steps. 

To analyze this problem,  we construct a controllability matrix associated with the automaton and the choice of control sites.  For simplicity, we restrict our attention to Boolean cellular automata, possibly defined in arbitrary dimension and allowing for possible inhomogeneities in their local update rules. We show that the system can be controlled only if this matrix is invertible, using Boolean operation \bassem{that is, if it has full Boolean rank. When this condition holds, its Boolean inverse gives a direct way to compute a control sequence that steers the system from any initial configuration to any desired final one.}

\bassem{The paper is organized as follows. Section~\ref{sec:definitions} introduces cellular automata, with emphasis on one- and two-dimensional models, and recalls adjacency matrices, Boolean derivatives, and the associated Jacobian matrix.  Section~\ref{sec:controlTheory} develops the control theory for linear one-dimensional elementary cellular automata. Section~\ref{sec:2DCA} extends the approach to two-dimensional linear CA and presents illustrative examples. Finally, Section~\ref{sec:conclusions} concludes the paper.}


\section{Definitions}\label{sec:definitions}
A cellular automaton is defined by a set of cells (sites), which can assume one out a finite set of states. There is an adjacency matrix that establishes the neighborhood of a cell, i.e., the set of cells whose state can determine the state of the site under examination. This dependency is given by an evolution function of the neighborhood. The evolution of the whole system is given by the simultaneous application of the evolution function to all cells.  

\subsection{Cellular automata}
Let $N$ denote the number of cells of the automaton, and 
let us denote by $@s\equiv @s(t)$ the configuration of the whole system at time $t$, with 
\eq[l1d]{
@s=(s_0, s_1, \dots, s_{N-1}).
}

The quantity $@s$ can be seen as an $N$-dimensional Boolean vector. We shall denote by $s_i$ the value of the $i$-th cell. We shall limit here to Boolean CA, for which $s_i\in\{0,1\}$, but our theory can be extended to other cases. 

The adjacency matrix $@a$ is such that $a_{ij}=1$ if cell $j$ belongs to the neighborhood of cell $i$ and zero otherwise. 

\begin{figure}[t]
    \centering
    \begin{tabular}{cc}
    (a) & (b) \\
    \includegraphics[width=0.45\linewidth]{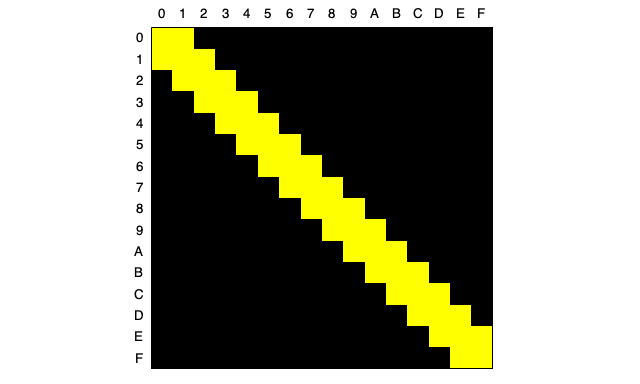} & 
    \includegraphics[width=0.45\linewidth]{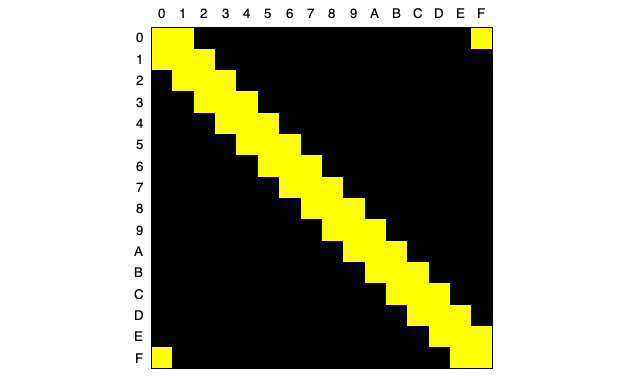}
    \end{tabular}
    \caption{The adjacency matrix of a 16-sites one-dimensional elementary cellular automaton of $16$ sites with (a) fixed or (b) periodic boundary conditions (indexed using the hexadecimal representation of indices). Yellow squares correspond to $a_{ij}=1$ and black squares to $a_{ij}=0$.}
    \label{fig:T16p}
\end{figure}

To be concrete, let us give some examples.
For one-dimensional elementary cellular automata, the neighborhood of site $i$ is composed by the cell itself and its nearest neighbors. Let us number the cells from left to right as in \Eq{l1d}.
The adjacency matrix for a lattice of $16$ sites with fixed and periodic boundary conditions is shown in Fig.~\ref{fig:T16p}.

For two-dimensional cellular automata of dimension $N=L\times L$, we identify cells by a sequential index,
\eq[l2d]{
    @s=\begin{pmatrix}
        s_0 & s_1 & s_2 & \dots & s_{L-1}\\
        s_L & s_{L+1} & s_{L+2} &\dots & s_{2L-1} \\
        \dots \\
        s_{L(L-1)} & s_{L(L-1)+1} & s_{L(L-1)+2} &\dots & s_{L^2-1}
    \end{pmatrix}.
}

\begin{figure}[t]
    \centering
    \begin{tabular}{cc}
    (a) & (b) \\
    \includegraphics[width=0.4\linewidth]{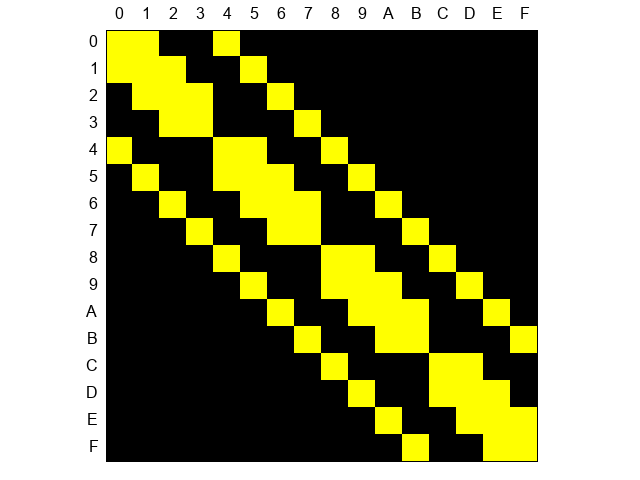}& 
    \includegraphics[width=0.4\linewidth]{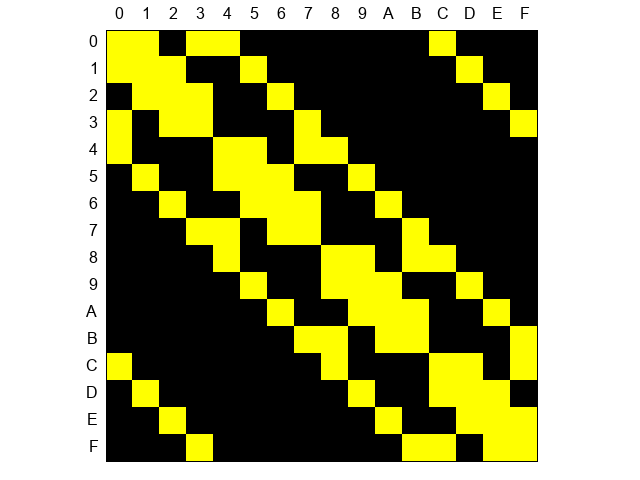}
    \end{tabular}
    \caption{The adjacency matrix of a two-dimensional cellular automaton of $4\times 4$ sites with von Neumann neighborhood and (a) fixed or (b) periodic boundary conditions. Yellow squares correspond to $a_{ij}=1$ and black squares to $a_{ij}=0$. }
    \label{fig:V4p}
\end{figure}

The adjacency matrix for a neighborhood that includes the cells itself and its nearest neighbors (von Neumann neighborhood), for a lattice with $L=4$ is reported in Fig.~\ref{fig:V4p}.

The evolution of the automaton is given by an $N$-dimensional Boolean function $F$ such that 
\eq{
    @s' = F(@s(t)),
}
where $@s'\equiv @s(t+1)$. 

In general, the global function $F$ is defined by means of a local function $f_i$, which may depend on the site index $i$ (inhomogeneous automata). For instance, for fixed boundary conditions, we can either impose that the sites at border do not evolve, or that the cells in the interior have a different neighborhood, in any case it is an inhomogeneous automaton. 

Let us denote by $@v_i$ the neighborhood of cell $i$, i.e., 
\eq{
    @v_i = (s_j : a_{ij}=1).
}
The argument of the function $f_i$ is the neighborhood $@v_i$ of the site $i$ under consideration.

The evolution of the lattice is given by the parallel application of local rule
\eq{
    s'_i = f_i(@v_i).
}

\subsection{Boolean derivatives}
Let us define the Boolean derivative~\cite{Vichniac1990} as 
\eq{
    \pdv{f(s_1,s_2,\dots,s_i, \dots s_n)}{s_i} = f(s_1,s_2,\dots,(s_i\oplus 1), \dots s_n)\oplus f(s_1,s_2,\dots,s_i, \dots s_n),
}
where $\oplus$ denotes the sum modulo two (the XOR Boolean function). 

The Boolean derivatives follows the usual rules, replacing the sum with the XOR and the multiplication with the AND (which behaves as the standard multiplication)~\cite{bagnoli1992boolean}.

As for standard derivative, we say that a function $f(x, y, \dots)$ depends linearly on the variable $x$ if $\partial f/\partial x = 1$. A linear function $f(x_1, x_2, \dots)$ is such that $\partial f/\partial x$ is constant, either zero or one. Notice that this definition of linear functions includes affine ones.

For instance, if $f(x, y, z) = x\oplus y \oplus 1$, then
\eq{
    \pdv{f(x, y)}{x}=\pdv{f(x, y)}{y}=1; \quad \pdv{f(x, y)}{z} = 0
}
(a linear rule), while for $f(x, y) = xy$
\eq{
    \pdv{f(x, y)}{x}=y \qquad \pdv{f(x, y)}{y}=x.
}

Finally, we can have functions that depends linearly only on some variables. For instance, for $f(x,y,z) = xy \oplus z\oplus 1$ we have
\eq{
      \pdv{f(x, y,z)}{x}=y \qquad \pdv{f(x,y,z)}{y}=x\qquad
  \pdv{f(x,y,z)}{z}=1,
}
which we call a right-peripherally linear rule.

Given the evolution rule 
\eq[F]{
    @s'=F(@s),
}
we can define the Jacobian 
\eq{
    J_{ij} = \pdv{s'_i}{s_j}=\pdv{f_i(@v_i)}{s_j},
}
which is a $N\times N$ Boolean matrix. 

The Jacobian $J_{ij}$ expresses the dependence of site $i$ at time $t+i$ with respect to a variation of site $j$ at time $t$, or, in other words, if $J_{ij}=1$ one can ``flip'' the value of $s_i$ at time $t+1$ by flipping the value of $s_j$ at time $t$. 

The Jacobian $@J$ is a constant matrix if all the local functions $f_i$ are linear, otherwise it is a function of the local configuration, $@J(t)\equiv @J(@s(t))$. The Jacobian $@J$ coincides with the adjacency matrix $a$ if all the functions $f_i(@v_i)$ depend linearly on all the variables in $@v_i$, otherwise $J_{ij} \le a_{ij}$.  

Due to the chain rule of Boolean derivatives~\cite{bagnoli1992boolean}, the derivatives at subsequent times are given by the Boolean product of Jacobians, where the sum is replaced by the XOR. We shall indicate the Boolean matrix product by $\odot$. Therefore, 
\eq{
    \pdv{s_i(T)}{s_j(0)} =\left[\bigodot_{t=1}^T @J(@s(t))\right]_{ij}
}
and in case of linear rules
\eq{
    \pdv{s_i(T)}{s_j(0)} =(@J^T)_{ij},
}
where the power of $@J$ is obtained by Boolean products.

\section{Control of cellular automata}\label{sec:controlTheory}
Let us now define the (ideal) control problem. Given a cellular automaton, the problem is that of choosing a minimal set $@c$ of $M$ control cells, and a minimal time horizon $T$, so that, for any initial configuration $@s_0=@s(0)$ of the rest of the system (i.e., the cells which are not chosen for control), and any target configuration $\tilde{@s}$, it is possible to find a sequence of states $(@c(t))_{t=0,\dots,T-1}$ for the control cells, such that the application of the rule of \Eq{F}, which we can rewrite as
\eq[CF]{
    @s(t+1) = F(@s(t), @c(t)),
}
is such that $@s(T)=\tilde{@s}$. 

The basic idea for checking is this is possible and to find the actual sequence is the following: let us choose an arbitrary setup of the states of the control sites, $@c^{(0)}$, for instance all zeros. We let the system evolve from $@s_0$ to $s(T)$, which in general is different from $\tilde{@s}$. Let us suppose that cell $s_i(T)$ has a state different from the desired one. If $c_j$ is a control site, and $\pdv*{s_i(T)}{c_j(t)}=1$ for some time step $0\let\<T$, then it is possible to ``flip'' the value of $s_i(T)$ by flipping that of $c_j(t)$. 

However, by flipping $c_j(t)$, it generally happens that other sites flip their values (i.e., $\pdv*{s_k(T)}{c_j(t)}=1$ for $k\neq i$). So we want to derive the conditions for which one can choose a combination of control values (to be imposed at specified time steps) so to flip any desired combination of sites in the target configuration $@s(T)$, given a starting configuration $@s(0)$, thus driving the system to the desired state. 

\subsection{Control of one-dimension cellular automata}

As  shown in Refs~\cite{bagnoli2016toward,dridi2022kalman}, control is always possible for one-dimensional cellular automata and peripherally-linear rules. In this case the Jacobian is a band-diagonal matrix with at least one constant external diagonal of ones. 

Let us consider for instance the case of an elementary CA (neighborhood formed by the cell itself and its nearest neighbors) in which there is a single control $c(t)$ acting on site $i=-1$, and fixed right boundary conditions, for instance $s_{N}(t)=0 \quad \forall t$. A symbolic evolution for $N=T=4$ is  
\eq[lc]{
    \begin{matrix}
        \textcolor {blue}{c(0)} & s_0(0) & s_1(0)& s_2(0)& s_3(0)& 0\\
        c(1) & \textcolor {blue}{s_0(1)} & s_1(1)& s_2(1)& s_3(1)&0\\
        c(2) & \textcolor {red}{s_0(2)} & \textcolor {blue}{s_1(2)}& s_2(2)& s_3(2)&0\\
        c(3) & \textcolor {red}{s_0(3)} & \textcolor {red}{s_1(3)}& \textcolor {blue}{s_2(3)}& s_3(3)&0\\
         & \textcolor {red}{s_0(4)} & \textcolor {red}{s_1(4)}& \textcolor {red}{s_2(4)}& \textcolor {blue}{s_3(4)}\\
    \end{matrix},
}
showing the propagation of a variation of the control for left-peripherally linear rule. 

Assume that $@s(0)=(s_0(0),\dots,s_3(0))$ is the starting configuration and $@s(4)=(s_0(4),\dots,s_3(4))$ the target one. If the value of $s_3(4)$ is not the desired one, one can flip  $c(0)$ and, due to the fact that leftmost nonzero diagonal of $@J$ only contains ones, it triggers the flip of $s_3(4)$ (blue chain in \Eq{lc}). This may change the values of red variables, but then one may act on the control at subsequent times to make the desired target configuration to appear. 

However, even for linear rules, this scheme cannot be applied to cellular automata with larger neighborhood (except for the special case of peripherally-linear CA and control placed at the border~\cite{bagnoli2016toward}, and higher-dimensional cellular automata. 

\subsection{General control theory for Boolean CA}

Let us use the index $i\in\{0,N-1\}$ for identifying the sites in $@s$ and the index $k$ for identifying the sites in the control set $@c$. 

Let us define the (Boolean) control matrix 
\eq{
    C_{in} \equiv C_{i,n(k,t)} =\pdv{s_i}{c_k(t)}
}
which expresses wether site $i$ at the  time step $T$ can be changed by flipping the value of control $k$ (among the $K$ possible ones) at time $t<T$. 

The index $n\equiv n(k,t)$  is obtained from $k\in (0,\dots,K-1)$ and $t\in (0,\dots,T-1)$ in order to identify which control is active at which time. We set 
\eq[n]{
    n = kT + t
}
so that each row $i$ of $@C_{in}$ is formed by $K$ blocks of length  $T$, each block determining the effects that the control $k$ (activated at different times) exerts on site $i$ at time $T$. 

We can see the problem as a set of Boolean equations
\eq{
    @\delta  @s(T) = @C @c
}
and the problem can be inverted 
\eq{
    @\delta  @c = @C^{-1} @\delta@s(T) = @A @\delta@s(T),
}
i.e., it is possible to find a sequence/choice of control values so to flip the desired sites in configuration $@\delta @s(T)$, if the rank of $@C$ (determined using Boolean operations) is $N$.
The matrix $@C^{-1}$ can be called the actuator matrix $@A$, since it indicates which control has to be turned on at which time step in order to achieve a given effect on a given cell at time $T$. 

Clearly, in order to control the $N$ cells of the control area we need at least the same quantity as the product of the number of controls sites and the number of time steps, but this is not sufficient, as we shall see. 

For linear rules, $@C$ is a constant matrix $N\times KT$, and can be constructed by the patterns generated by the effects of controls at time 0. 

Let us for instance obtain the control matrix $C$ for a one-dimensional, 4-sites linear elementary CA ($f(x,y,z)=x\oplus y\oplus z$) with fixed boundaries (geometry as in \Eq{lc}), whose Jacobian is 
\eq{ 
@J = \begin{bmatrix}
    1 & 1 & 0 & 0 \\
    1 & 1 & 1 & 0 \\
    0 & 1 & 1 & 1 \\
    0 & 0 & 1 & 1 \\
    \end{bmatrix},
}
with a single ($k=0$) control at left (replacing the fixed left boundary). 

The effect of the control $c(0)=c_0(0)$ on the configuration  $@s$ is that of flipping site $i=0$ at time $t=1$, and can be represented by the vector 
\eq{
    @{\delta}@s(1) = (1,0,0,0)^\dagger,
}
where the symbol $\dagger$ denotes the transpose. 

This is also the effect of  $c(T-1)$, i.e., the control applied at time $T-1$, on the configuration $@s(T)$ at time $T$, so $C_{i,T-1} = (@{\delta}@s(1))_i$, $i=1,\dots,4$. 

If the same control is applied at time $T-2$ (and no other controls after it), its effects are given by 
\eq{
    @{\delta}@s(2) = @J @{\delta}@s(1),
}
which thus gives $C_{i,T-2}=(@J@{\delta}@s(1))_i$.

Therefore, we can obtain the $k$-th block of the control matrix by setting the column $kT-1$ of $C$ equal to the the effects (modifications) of control $k$ on the configuration at the following time step (i.e., control $k$ at time $T-1$ on $s(T)$), and then iterating ``backward'', for each block $k$, by applying the Jacobian:
\eq{
    C_{(k-1)T+t-1,j} = \bigoplus_{j=1}^{N-1} J_{ij} C_{j, (k-1)T+t}, \quad\text{for}\quad t=T-1,\dots, 1.
}

For the previous example with $N=T=4$, and a single control $k=0$ we have
\eq{
    \begin{pmatrix}
        \delta s_0(4)\\
        \delta s_1(4)\\
        \delta s_2(4)\\
        \delta s_3(4)\\
    \end{pmatrix} = @C 
    \begin{pmatrix}
        c(0)\\
        c(1)\\
        c(2)\\
        c(3)\\
    \end{pmatrix}
    =
    \begin{pmatrix}
       0 & 0 & 1 & 1\\
        1 & 0 & 1 & 0 \\
        1 & 1 & 0 & 0 \\
        1 & 0 & 0 & 0
    \end{pmatrix}
    \begin{pmatrix}
        c(0)\\
        c(1)\\
        c(2)\\
        c(3)\\
    \end{pmatrix},
}
and we can check that the matrix $@C$ can be inverted, which is consistent to the fact that the rule is peripherally linear~\cite{bagnoli2016toward,dridi2022kalman,dridi2019recent}, 
giving 
\eq{
        \begin{pmatrix}
        c(0)\\
        c(1)\\
        c(2)\\
        c(3)\\
    \end{pmatrix} = @A 
     \begin{pmatrix}
        \delta s_0(4)\\
        \delta s_1(4)\\
        \delta s_2(4)\\
        \delta s_3(4)\\
    \end{pmatrix}
    =
    \begin{pmatrix}
     0  &   0  &   0  &   1\\
     0   &  0   &  1   &  1\\
     0   &  1   &  0  &   1\\
     1   &  1  &   0   &  1
    \end{pmatrix} 
    \begin{pmatrix}
        \delta s_0(4)\\
        \delta s_1(4)\\
        \delta s_2(4)\\
        \delta s_3(4)\\
    \end{pmatrix},
}
indicating that, for instance, if one wants to flip site $0$ ($@\delta@s(4)=(1,0,0,0)^\dagger$), one has to apply the control at time 3, $@c=(0,0,0,1)^\dagger$ (the $\dagger$ symbol indicate the transpose), but if one wants to flip site $3$, one has to apply controls at times $0,1,2,3$, and to flip simultaneously sites 1 and 3, one has to apply controls  at times $0$ and $1$ (all sum operations are done modulo 2, so that in this last case the sum of contributions in positions 2 and 3 cancels out).

In order to determine the rank of the control matrix and to invert it (or at least find the maximum number of independent sites that can be controlled), one has to use methods compatible with the Boolean character of the matrix, using only Boolean operations. One possibility is to use a variation of the Gauss-Jordan procedure: 
\begin{enumerate}
\item  Augment the control matrix $@C$ by an identity $N\times N$ matrix $I$: $@A = (@C,@I)$.
\item Reduce the matrix $A$ to a diagonal form by summing (modulo two) the rows or swapping them if the leading value is zero. 
\item The number of non-zero diagonals give the independent controls.
\item The transformed identity matrix furnishes the inverted control matrix.
\end{enumerate}

For example, let us use the same example as before, but apply the control at left ($k=0$) and at right ($k=1$) for two time steps,
\eq{
    \begin{matrix}
        c_0(0) & s_0(0) & s_1(0) & s_2(0)& s_3(0) &c_1(0)\\
        c_0(1) & s_0(1) & s_1(1) & s_2(1)& s_3(1)&c_1(1)\\
         & s_0(2) & s_1(2)& s_2(2)& s_3(2)&\\
    \end{matrix}\quad.
}

We get 
\eq{
        \begin{pmatrix}
        \delta s_0(2)\\
        \delta s_1(2)\\
        \delta s_2(2)\\
        \delta s_3(2)\\
        \end{pmatrix}
        =@C
        \begin{pmatrix}
        c_0(0)\\
        c_0(1)\\
        c_1(0)\\
        c_1(1)\\
        \end{pmatrix}
     = \begin{pmatrix}
        1 & 1 & 0 & 0\\
        1 & 0 & 0 & 0\\
        0 & 0 & 1 & 0\\
        0 & 0 & 1 & 1\\
        \end{pmatrix} 
        \begin{pmatrix}
        c_0(0)\\
        c_0(1)\\
        c_1(0)\\
        c_1(1)\\
        \end{pmatrix},
}
and inverting the matrix $@C$  one obtains
\eq{
    \begin{pmatrix}
        c_0(0)\\
        c_0(1)\\
        c_1(0)\\
        c_1(1)\\
    \end{pmatrix}= 
    @C^{-1} 
    \begin{pmatrix}
        s_0(4)\\
        s_1(4)\\
        s_2(4)\\
        s_3(4)\\
     \end{pmatrix} =
    \begin{pmatrix}
    0 & 1 & 0 & 0\\
    1 & 1 & 0 & 0\\
    0 & 0 & 1 & 0 \\
    0 & 0 & 1 & 1 \\
    \end{pmatrix}
    \begin{pmatrix}
        s_0(4)\\
        s_1(4)\\
        s_2(4)\\
        s_3(4)\\
     \end{pmatrix}. 
 }

Let us now present the cases of  over-controlled and under-controlled situations. 
For the first case, let us consider the same problem with right and left controls, but for $T=3$,
\eq{
    \begin{matrix}
        c_0(0) & s_0(0) & s_1(0) & s_2(0)& s_3(0) &c_1(0)\\
        c_0(1) & s_0(1) & s_1(1) & s_2(1)& s_3(1)&c_1(1)\\
        c_0(2) & s_0(2) & s_1(2)& s_2(2)& s_3(2)&c_1(2)\\
         & s_0(3) & s_1(3) & s_2(3)& s_3(3) &
    \end{matrix}\quad.
}
The control matrix $@C$ is 
\eq{
    @C=\begin{pmatrix}
        0 & 1 & 1 & 0 & 0 & 0 \\
        0 & 1 & 0 & 1 & 0 & 0 \\
        1 & 0 & 0 & 0 & 1 & 0 \\
        0 & 0 & 0 & 0 & 1 & 1 
    \end{pmatrix}
}
and in this case the Gauss-Jordan diagonalization has to applied by columns, or to the transpose matrix, obtaining (eventually after another transposition),
\eq{
    \begin{pmatrix}
        c_0(0)\\
        c_0(1)\\
        c_0(2)\\
        c_1(0)\\
        c_1(1)\\
        c_1(2)\\
    \end{pmatrix}
    = \begin{pmatrix}
    0 & 0 & 1 & 0 \\
    0 & 1 & 0 & 0\\
    1 & 1 & 0 & 0 \\
    0 & 0 & 0 & 0 \\
    0 & 0 & 0 & 0 \\
    0 & 0 & 0 & 1\\
    \end{pmatrix}
    \begin{pmatrix}
        \delta s_0(4)\\
        \delta s_1(4)\\
        \delta s_2(4)\\
        \delta s_3(4)\\
     \end{pmatrix} 
 }    
where $c_1(0)$ and $c_1(1)$ are not used.
Since the system is over-constrained, one could obtain a different control scheme if the Gauss-Jordan procedure is performed in a  different order, for instance not using controls at time 0,  reverting to the previous case.

\begin{figure}[t]
    \centering
    \begin{tabular}{cccc}
    (a) & (b)  & (c) & (d) \\
    \includegraphics[width=0.2\linewidth]{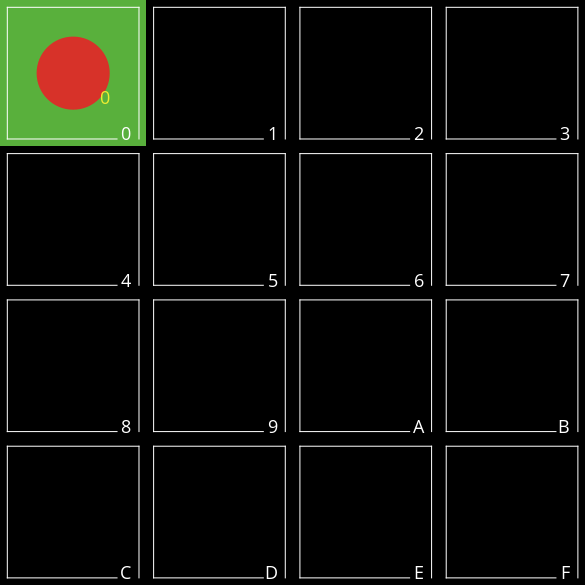} &
    \includegraphics[width=0.2\linewidth]{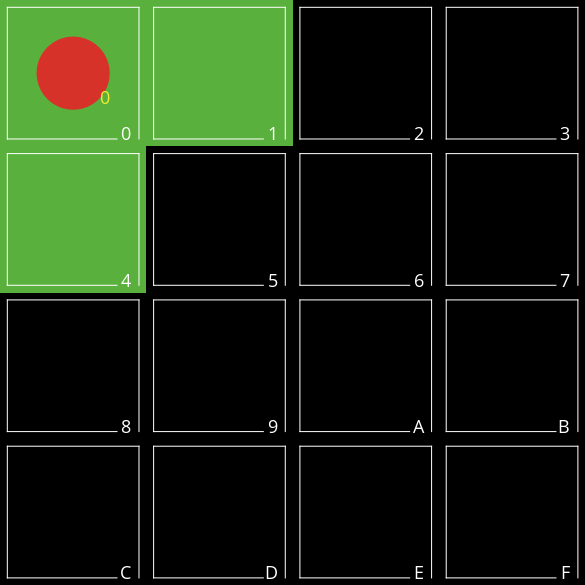} &
    \includegraphics[width=0.2\linewidth]{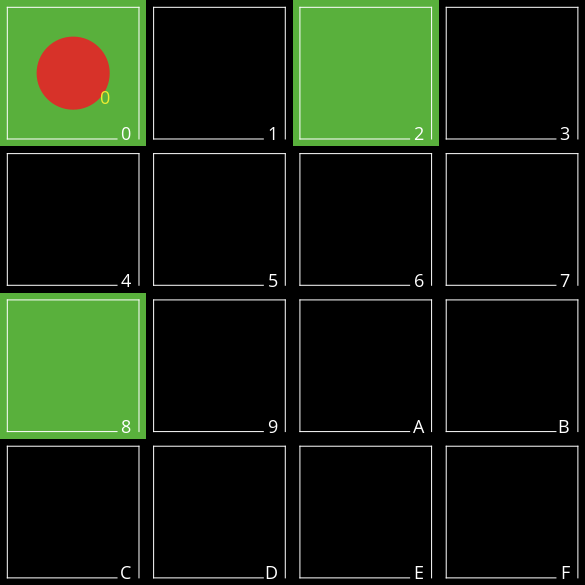} &
    \includegraphics[width=0.2\linewidth]{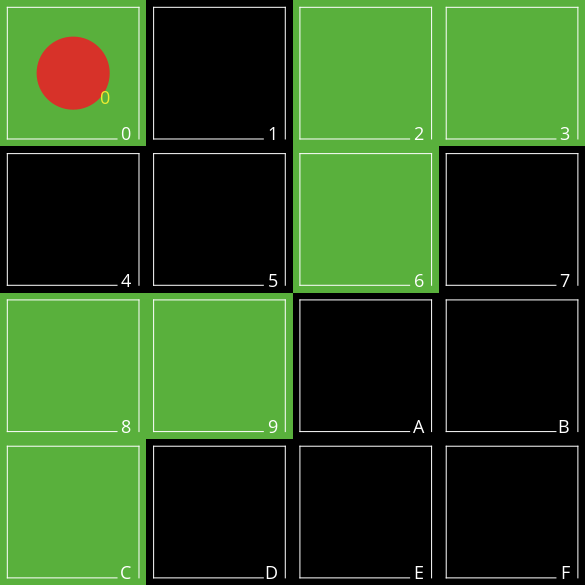} \\
    \end{tabular}
    \caption{The evolution of an initial perturbation at time $t=0$ on site $(0,0)$ (red circle) for a linear rule and von Neumann neighborhood, zero boundary conditions. (a) $t=1$, (b) $t=2$, (c) $t=3$, (d) $t=4$, after which the patterns (c) and (d) repeat themselves. Color code: black=0, green = 1.}
    \label{fig:VN1}
\end{figure}

For the under-constrained problem, we take the same example but with only left controls,
\eq{
    \begin{matrix}
        c_0(0) & s_0(0) & s_1(0) & s_2(0)& s_3(0) &0\\
        c_0(1) & s_0(1) & s_1(1) & s_2(1)& s_3(1)&0\\
        c_0(2) & s_0(2) & s_1(2)& s_2(2)& s_3(2)&0\\
         & s_0(3) & s_1(3) & s_2(3)& s_3(3) &
    \end{matrix}\quad.
}
The control matrix is 
\eq{
    @C=\begin{pmatrix}
        0 & 1 & 1  \\
        0 & 1 & 0  \\
        1 & 0 & 0  \\
        0 & 0 & 0  
    \end{pmatrix}
}
and we get 
\eq{
    \begin{pmatrix}
        c_0(0)\\
        c_0(1)\\
        c_0(2)
    \end{pmatrix}
    = \begin{pmatrix}
    0 & 0 & 1 & 0\\
    0 & 1 & 0 & 0\\
    1 & 1 & 0 & 0\\
    \end{pmatrix}
    \begin{pmatrix}
        \delta s_0(3)\\
        \delta s_1(3)\\
        \delta s_2(3)\\
        \delta s_3(3)\\
     \end{pmatrix},
 }  
where $s_3(3)$ cannot be controlled. 

\section{Two-dimensional cellular automata}
\label{sec:2DCA}

\begin{figure}[t]
    \centering
    \begin{tabular}{cccc}
    (a) & (b)  &  (c) & (d)  \\
    \includegraphics[width=0.2\linewidth]{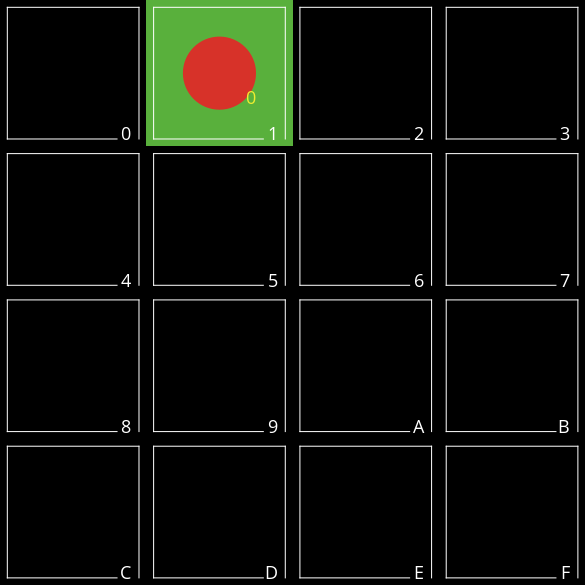} &
    \includegraphics[width=0.2\linewidth]{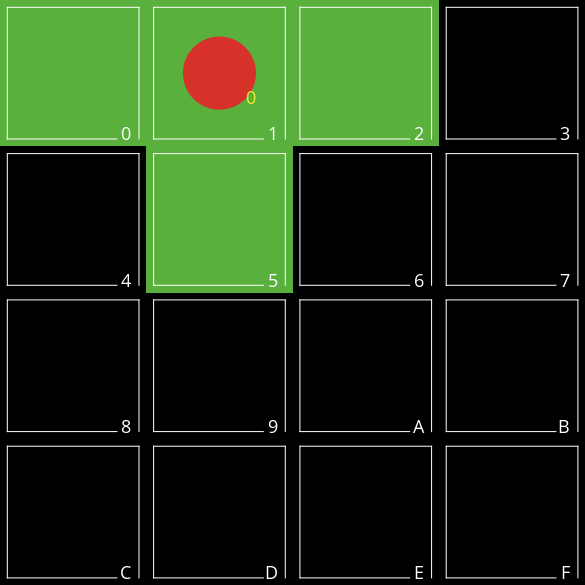} &
    \includegraphics[width=0.2\linewidth]{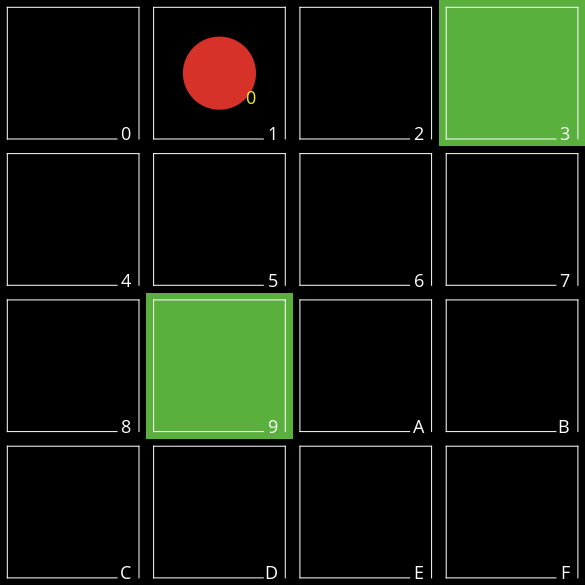} &
    \includegraphics[width=0.2\linewidth]{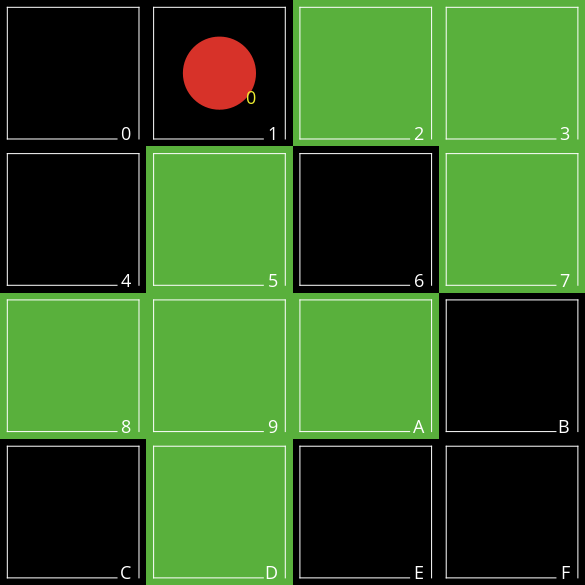} \\
    \end{tabular}
    \caption{The evolution of an initial perturbation at $t=0$ on site $(0,1)$ (red circle) for a linear rule and von Neumann neighborhood, zero boundary conditions. (a) $t=1$, (b) $t=2$, (c) $t=3$, (d) $t=4$, after which the patterns (c) and (d) repeat themselves. Color code: black=0, green = 1.}
    \label{fig:VN2}
\end{figure}

Let us investigate the control problem for a 2-dimensional linear rule with von Neumann neighborhood, fixed boundary conditions (Fig.~\ref{fig:V4p}-(a)) and a control exerted at time $t=0$ on site $(0,0)$ (sequential site 0, index $n= 0$, \Eq{n}) and $(0,1)$ (sequential site 1, index $n=1$). Their effects are  shown in Figs.~\ref{fig:VN1} and \ref{fig:VN2}. 

One can see that the perturbation never reaches some sites: 
in all cases there are 6 sites that are insensitive to any control (due to interference effects), but there are also other sites that receive the same sequence of controls, so that they cannot be controlled independently, although the rule is linear. It is therefore not clear if the system can be controlled and, moreover, which is the sequence of controls needed to modify an existing configuration. 

Since the effects of a control on a boundary are effective only for 4 time steps, and we need 16 independent control combination, we can try with 4 controls (sites 0,1,2,3) for 4 time steps.
The corresponding control matrix is 
\eq{
\scriptstyle
\begin{pmatrix}
\delta s_0(4)\\
\delta s_1(4)\\
\delta s_2(4)\\
\delta s_3(4)\\
\delta s_4(4)\\
\delta s_5(4)\\
\delta s_6(4)\\
\delta s_7(4)\\
\delta s_8(4)\\
\delta s_9(4)\\
\delta s_\A(4)\\
\delta s_\B(4)\\
\delta s_\C(4)\\
\delta s_\D(4)\\
\delta s_\E(4)\\
\delta s_\F(4)\\
\end{pmatrix} 
= 
\begin{pmatrix}
1 & 1 & 1 & 1 &| & 0 & 0 & 1 & 0 & | & 1 & 1 & 0 & 0 &| & 1 & 0 & 0 & 0 \\
0 & 0 & 1 & 0 &| & 0 & 0 & 1 & 1 & | & 1 & 0 & 1 & 0 &| & 1 & 1 & 0 & 0 \\  
1 & 1 & 0 & 0 &| & 1 & 0 & 1 & 0 & | & 0 & 0 & 1 & 1 &| & 0 & 0 & 1 & 0 \\
1 & 0 & 0 & 0 &| & 1 & 1 & 0 & 0 & | & 0 & 0 & 1 & 0 &| & 1 & 1 & 1 & 1 \\
0 & 0 & 1 & 0 &| & 0 & 0 & 0 & 0 & | & 1 & 0 & 0 & 0 &| & 0 & 0 & 0 & 0 \\
0 & 0 & 0 & 0 &| & 1 & 0 & 1 & 0 & | & 0 & 0 & 0 & 0 &| & 1 & 0 & 0 & 0 \\
1 & 0 & 0 & 0 &| & 0 & 0 & 0 & 0 & | & 1 & 0 & 1 & 0 &| & 0 & 0 & 0 & 0 \\
0 & 0 & 0 & 0 &| & 1 & 0 & 0 & 0 & | & 0 & 0 & 0 & 0 &| & 0 & 0 & 1 & 0 \\
1 & 1 & 0 & 0 &| & 1 & 0 & 0 & 0 & | & 0 & 0 & 0 & 0 &| & 0 & 0 & 0 & 0 \\
1 & 0 & 0 & 0 &| & 1 & 1 & 0 & 0 & | & 1 & 0 & 0 & 0 &| & 0 & 0 & 0 & 0 \\
0 & 0 & 0 & 0 &| & 1 & 0 & 0 & 0 & | & 1 & 1 & 0 & 0 &| & 1 & 0 & 0 & 0 \\
0 & 0 & 0 & 0 &| & 0 & 0 & 0 & 0 & | & 1 & 0 & 0 & 0 &| & 1 & 1 & 0 & 0 \\
1 & 0 & 0 & 0 &| & 0 & 0 & 0 & 0 & | & 0 & 0 & 0 & 0 &| & 0 & 0 & 0 & 0 \\
0 & 0 & 0 & 0 &| & 1 & 0 & 0 & 0 & | & 0 & 0 & 0 & 0 &| & 0 & 0 & 0 & 0 \\
0 & 0 & 0 & 0 &| & 0 & 0 & 0 & 0 & | & 1 & 0 & 0 & 0 &| & 0 & 0 & 0 & 0 \\
0 & 0 & 0 & 0 &| & 0 & 0 & 0 & 0 & | & 0 & 0 & 0 & 0 &| & 1 & 0 & 0 & 0 \\
\end{pmatrix}
\begin{pmatrix}
c_0(0)\\
c_0(1)\\
c_0(2)\\
c_0(3)\\
c_1(0)\\
c_1(1)\\
c_1(2)\\
c_1(3)\\
c_2(0)\\
c_2(1)\\
c_2(2)\\
c_2(3)\\
c_3(0)\\
c_3(1)\\
c_3(2)\\
c_3(3)\\
\end{pmatrix} 
}
where we put into evidence the blocks corresponding to the different controls (from left to right $k=0,1,2,3$). We see here how the matric $@C$ is built in a complex case: each block $k$ is obtained taking for the last column of the block the effect of the control $k$ at time $T-1$ (column 3 for control $c_0(3)$, column 7 for control $c_1(3)$, column 11 for $c_2(3)$ and column 15 for $c_3(3)$) and then iteratively applying the Jacobian (equal here to the adjacency matrix) in order to obtain the effects of the same control applied at times $T-2,T-3,\dots$. 

\begin{figure}[t]
    \centering
    \begin{tabular}{cccc}
    \includegraphics[width=0.2\linewidth]{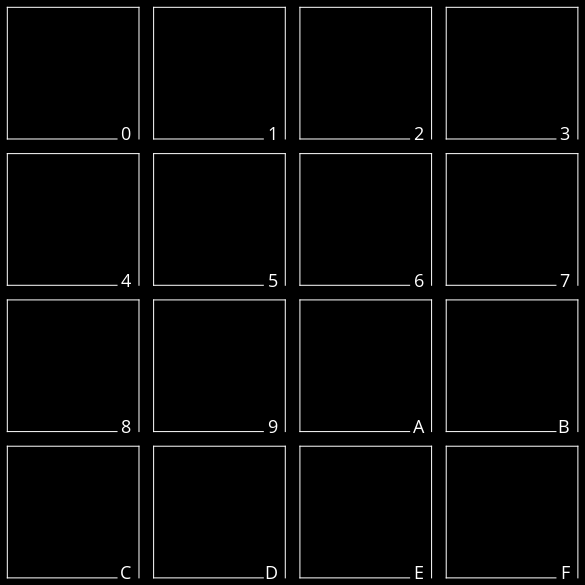} &
    \includegraphics[width=0.2\linewidth]{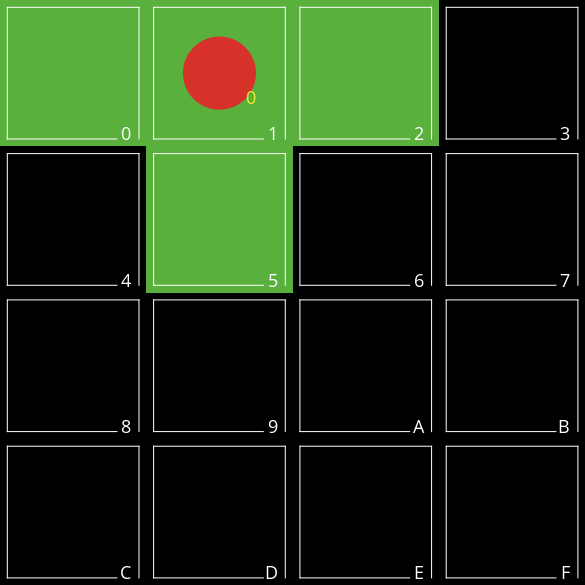} &
    \includegraphics[width=0.2\linewidth]{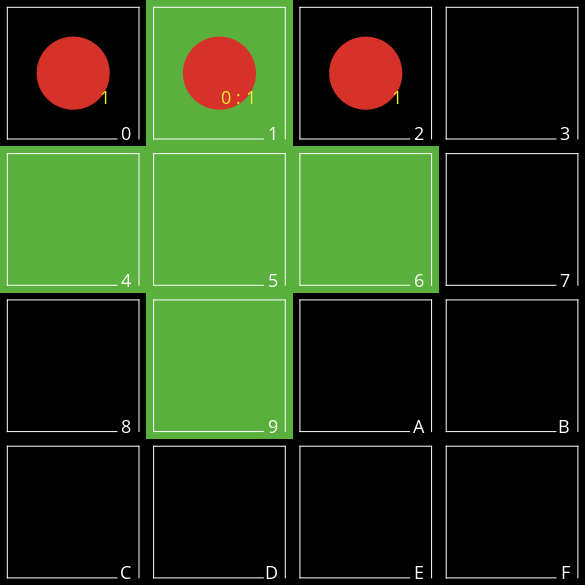} &
     \includegraphics[width=0.2\linewidth]{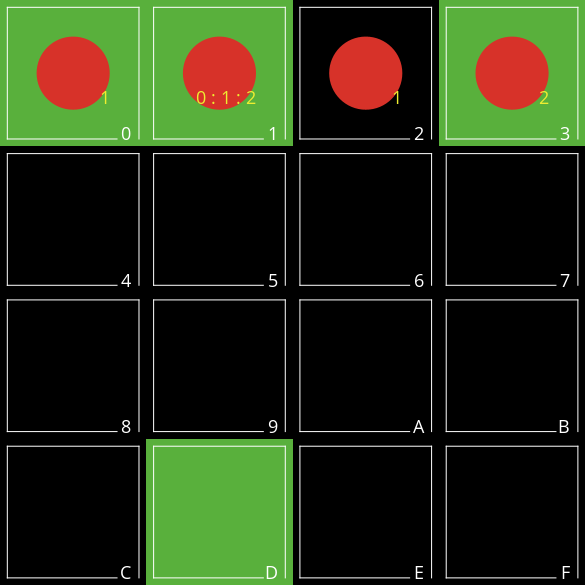} \\
   \includegraphics[width=0.2\linewidth]{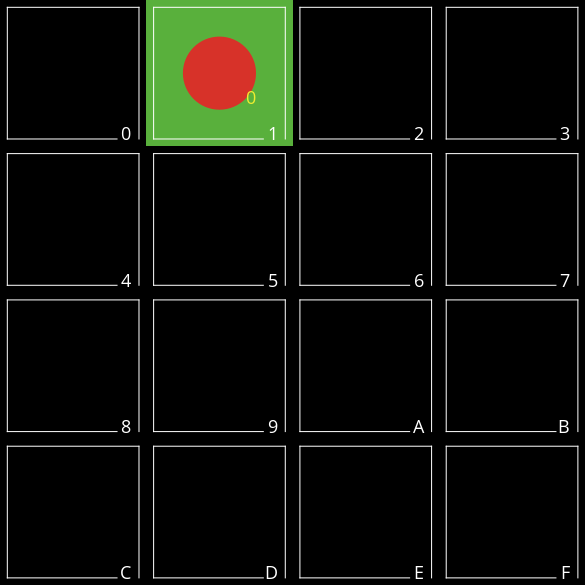} &
    \includegraphics[width=0.2\linewidth]{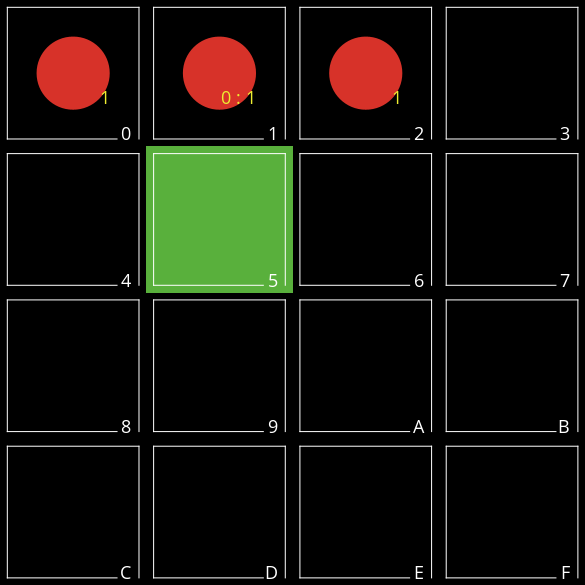} &
    \includegraphics[width=0.2\linewidth]{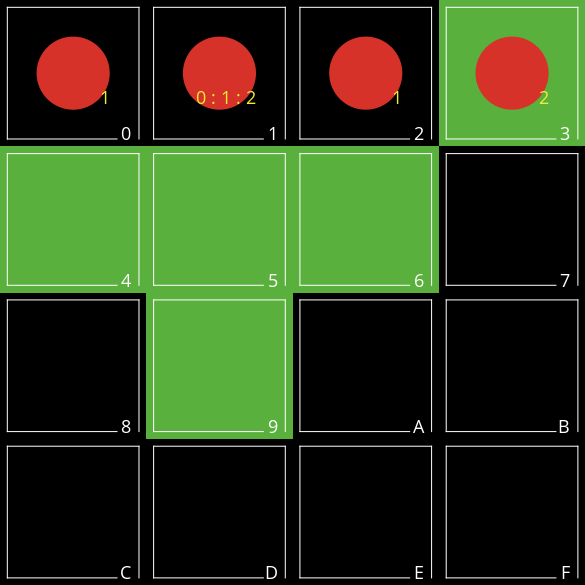} &
    \includegraphics[width=0.2\linewidth]{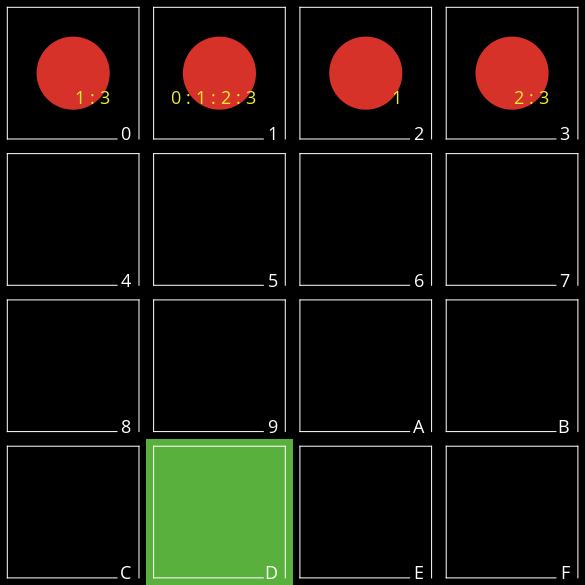} 
    \end{tabular}
    \caption{An example of application of the control in the two-dimensional Von Neumann linear rule with fixed boundaries, where we want to flip cell $D$ (13) at time $T=3$. top row: the configuration before applying the control, bottom row: after applying the control, i.e., with cell state flipped. Time is from left to right. The activation times of controls are marked in yellow.}
    \label{fig:C13}
\end{figure}

The corresponding actuator matrix is 
\eq[VNC]{
\begin{pmatrix}
c_0(0)\\
c_0(1)\\
c_0(2)\\
c_0(3)\\
c_1(0)\\
c_1(1)\\
c_1(2)\\
c_1(3)\\
c_2(0)\\
c_2(1)\\
c_2(2)\\
c_2(3)\\
c_3(0)\\
c_3(1)\\
c_3(2)\\
c_3(3)\\
\end{pmatrix} 
= 
\begin{pmatrix}
0 & 0 & 0 & 0 & 0 & 0 & 0 & 0 & 0 & 0 & 0 & 0 & 1 & \textcolor{red}{0} & 0 & 0 \\
0 & 0 & 0 & 0 & 0 & 0 & 0 & 0 & 1 & 0 & 0 & 0 & 1 & \textcolor{red}{1} & 0 & 0 \\
0 & 0 & 0 & 0 & 1 & 0 & 0 & 0 & 0 & 0 & 0 & 0 & 0 & \textcolor{red}{0} & 1 & 0 \\
1 & 0 & 0 & 0 & 1 & 1 & 0 & 0 & 1 & 0 & 1 & 0 & 0 & \textcolor{red}{1} & 1 & 1 \\
0 & 0 & 0 & 0 & 0 & 0 & 0 & 0 & 0 & 0 & 0 & 0 & 0 & \textcolor{red}{1} & 0 & 0 \\
0 & 0 & 0 & 0 & 0 & 0 & 0 & 0 & 0 & 1 & 0 & 0 & 1 & \textcolor{red}{1} & 1 & 0 \\
0 & 0 & 0 & 0 & 0 & 1 & 0 & 0 & 0 & 0 & 0 & 0 & 0 & \textcolor{red}{1} & 0 & 1 \\
0 & 1 & 0 & 0 & 1 & 1 & 1 & 0 & 0 & 0 & 0 & 1 & 1 & \textcolor{red}{1} & 0 & 1 \\
0 & 0 & 0 & 0 & 0 & 0 & 0 & 0 & 0 & 0 & 0 & 0 & 0 & \textcolor{red}{0} & 1 & 0 \\
0 & 0 & 0 & 0 & 0 & 0 & 0 & 0 & 0 & 0 & 1 & 0 & 0 & \textcolor{red}{1} & 1 & 1 \\
0 & 0 & 0 & 0 & 0 & 0 & 1 & 0 & 0 & 0 & 0 & 0 & 1 & \textcolor{red}{0} & 1 & 0 \\
0 & 0 & 1 & 0 & 0 & 1 & 1 & 1 & 1 & 0 & 0 & 0 & 1 & \textcolor{red}{0} & 1 & 1 \\
0 & 0 & 0 & 0 & 0 & 0 & 0 & 0 & 0 & 0 & 0 & 0 & 0 & \textcolor{red}{0} & 0 & 1 \\
0 & 0 & 0 & 0 & 0 & 0 & 0 & 0 & 0 & 0 & 0 & 1 & 0 & \textcolor{red}{0} & 1 & 1 \\
0 & 0 & 0 & 0 & 0 & 0 & 0 & 1 & 0 & 0 & 0 & 0 & 0 & \textcolor{red}{1} & 0 & 0 \\
0 & 0 & 0 & 1 & 0 & 0 & 1 & 1 & 0 & 1 & 0 & 1 & 1 & \textcolor{red}{1} & 1 & 0 \\ 
\end{pmatrix}
\begin{pmatrix}
\delta s_0(4)\\
\delta s_1(4)\\
\delta s_2(4)\\
\delta s_3(4)\\
\delta s_4(4)\\
\delta s_5(4)\\
\delta s_6(4)\\
\delta s_7(4)\\
\delta s_8(4)\\
\delta s_9(4)\\
\delta s_\A(4)\\
\delta s_\B(4)\\
\delta s_\C(4)\\
\textcolor{red}{\delta s_\D(4)}\\
\delta s_\E(4)\\
\delta s_\F(4)\\
\end{pmatrix} 
}
where we marked in red the control sequences needed to flip site $\D$ (13) at time $T=3$ (or to turn on only site $\D$ at time $T=3$ starting from a zero configuration, since the rule is linear), corresponding to Fig.~\ref{fig:C13}. 

\begin{figure}[t]
    \centering
    \begin{tabular}{cccc}
    (a) & (b) & (c) & (d) \\
    \includegraphics[width=0.2\linewidth]{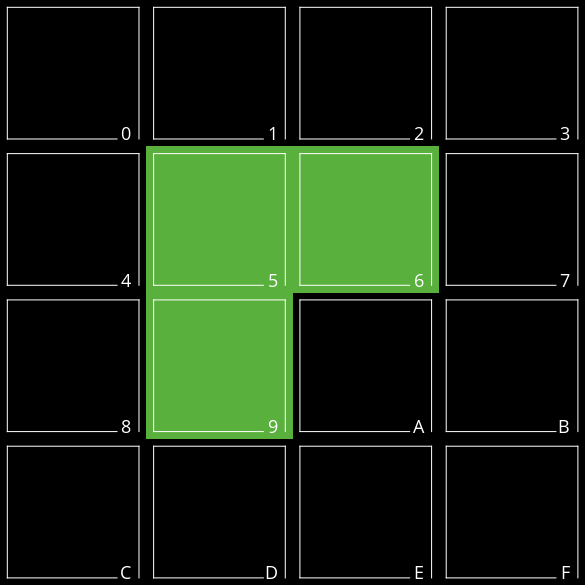} &
 \includegraphics[width=0.2\linewidth]{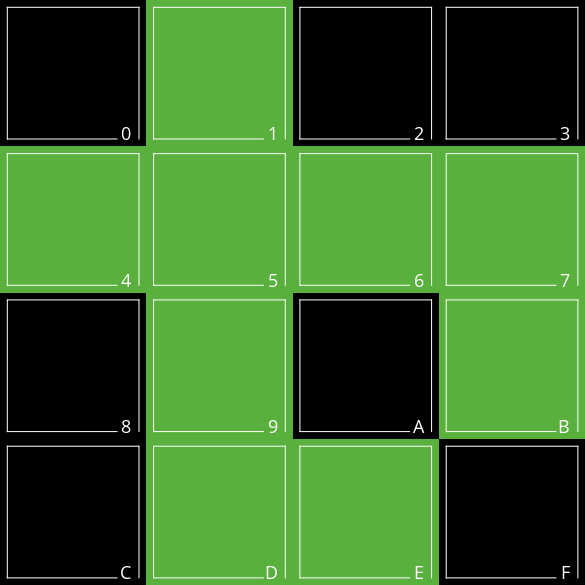}&
 \includegraphics[width=0.2\linewidth]{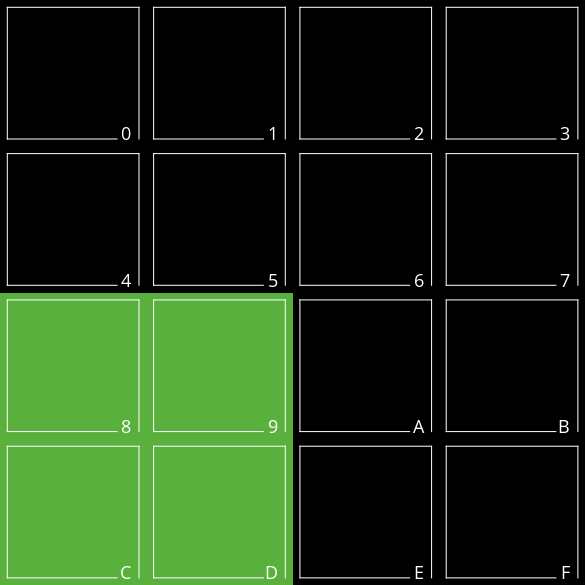} &
 \includegraphics[width=0.2\linewidth]{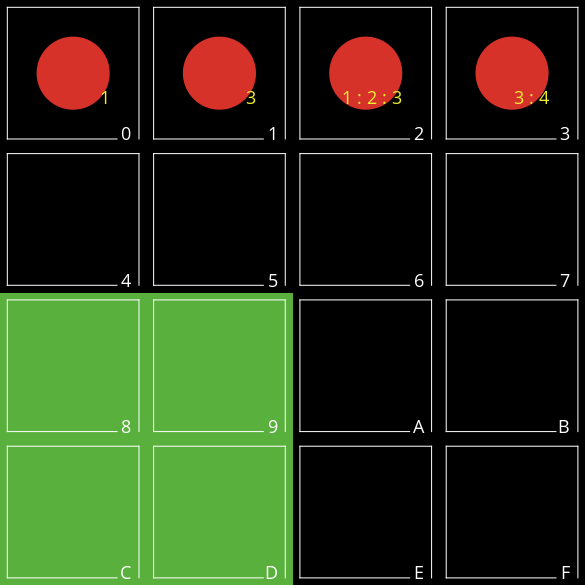}
 \end{tabular}
   \caption{(a) The initial configuration at $T=0$. (b) The configuration at time $T=4$ without controls. (c) The desired configuration. (d) The control sites (red circles) with the corresponding activation times in yellow. }
    \label{fig:VNT}
\end{figure}

And finally a real control experiment. Let us start from configuration with sites 5,6 and 9 on, Fig~\ref{fig:VNT}-(a), which, after four time steps generates the pattern of Fig~\ref{fig:VNT}-(b).  We however assume that the required pattern is that of Fig.~\ref{fig:VNT}-(c). 

We have therefore to flip sites $(1,4,5,6,7, 8,\B,\C,\E)$ (difference between Fig~\ref{fig:VNT}-(b) ad Fig~\ref{fig:VNT}-(c)). Using the corresponding vector $@\delta@s = (0,1,0,0,1,1,1,1,1,$ $0,0,1,1,0,1,0)^\dagger$ in \Eq{VNC}, we get the control sequence $@c=(1,0,0,0,0,0,1,0,1,1,$ $1,0,0,0,1,1)^\dagger$, also reported in Table.~\ref{tab:VNcontrols}, which indeed gives the desired pattern, as shown in Fig.~\ref{fig:VNT}-(d).

\begin{table}[t]
\centering
\begin{tabular}{c|cccc}
     & \multicolumn{4}{c}{sites}\\
 time  &     0 & 1 & 2 & 3 \\
  \hline
 0 &     1   &  0   &  1   &  0\\
 1 &    0   &  0   &  1    & 0 \\
 2 &     0  &   1   &  1   & 1 \\
 3 &    0  &   0   &  0    & 1
\end{tabular}
\caption{The sequence of controls reported in Fig.\ref{fig:VNT}-(d).} \label{tab:VNcontrols}
\end{table}

By complete enumeration, one finds that, for the $4\times 4$ lattice and $4$ time steps, the dispositions of 4 controls  that produce an invertible control matrix are not many, just 224 over 1820 possible ones (many of which are symmetric under rotation or mirroring). 

This makes this derivation more important, since it is already an hard task that of finding the valid positions of controls, not to say their order of activation to reach the desired target.

\section{Conclusions}\label{sec:conclusions}

In this paper, we tackled  the problem of finding a general control theory of linear cellular automata. We investigated here Boolean automata. By constructing an appropriate control matrix associated with the system dynamics, we established the necessary and sufficient conditions for controllability. We proved that the system is controllable if and only if the corresponding control matrix has full rank, i.e., it is invertible (using Boolean operations). 
Several illustrative examples are provided to demonstrate the applicability of the proposed approach in both one and two-dimensional cases.

\printbibliography
\end{document}